\documentclass[msmath,superscriptaddress,twocolumn,amssymb,prb]{revtex4}
\usepackage{graphicx}
\usepackage{dcolumn}
\usepackage{pdfsync}
\usepackage{bm}
\usepackage{hyperref}

\begin{document}
\title{{\sc Muse}: Multi-algorithm collaborative crystal structure prediction}
\author{Zhong-Li Liu}

\email{zl.liu@163.com}

\affiliation{College of Physics and Electric Information, Luoyang Normal University, Luoyang 471022, China}

\date{\today}
\begin{abstract}
The algorithm and testing of the Multi-algorithm-collaborative Universal Structure-prediction Environment ({\sc Muse}) are detailed. Presently, in {\sc Muse} I combined the evolutionary, the simulated annealing, and the basin hopping algorithms to realize high-efficiency structure predictions of materials under certain conditions. {\sc Muse} is kept open and other algorithms can be added in future. I introduced two new operators, slip and twist, to increase the diversity of structures. In order to realize the self-adaptive evolution of structures, I also introduced the competition scheme among the ten variation operators, as is proved to further increase the diversity of structures. The symmetry constraints in the first generation, the multi-algorithm collaboration, the ten variation operators, and the self-adaptive scheme are all key to enhancing the performance of {\sc Muse}. To study the search ability of {\sc Muse}, I performed extensive tests on different systems, including the metallic, covalent, and ionic systems. All these present tests show {\sc Muse} has very high efficiency and 100\% success rate.

\end{abstract}

%

\maketitle
\section{Introduction}
The structural data of materials are of fundamental importance to our understanding their properties. Usually, the structures are determined from experiments, such as the X-ray powder diffraction experiment. However, when materials are in very complex or extreme conditions, such as high pressure and/or high temperature~\cite{Oganov2006-3}, or their compositions are beyond our chemical intuition,~\cite{Lonie2011} experiment is often difficult or even impossible to identify their structures. However, from the theoretical aspect, the crystal structure prediction (CSP) methods are often efficient, economic, and convenient to determine the structures of materials.

Maddox claimed the ``scandal in the physical sciences" of the inability to predict the structure of a material from its stoichiometry two decades ago.~\cite{Maddox1988} Nevertheless, many groups have made great success in the CSP based on different stochastic optimization algorithms. Currently, the mature CSP methods are the metadynamics,~\cite{Martonak2003,Laio2002} the minima hopping,~\cite{Wales1997,Goedecker2004,Amsler2010} the evolutionary algorithm (EA),~\cite{Oganov2006-3,Lonie2011,Oganov2010,evo2013} the particle swarm optimization (PSO),~\cite{Wang2010,Wang2012,Lv2012} and the random search technique.~\cite{Pickard2011} Using these methods, many groups have discovered plenty of novel materials under ambient and extreme conditions. The post-perovskite phase of MgSiO$_3$,~\cite{Oganov2005} the new structure of P,~\cite{Ishikawa2006} and so on, were found using the metadynamics technique. Z-carbon,~\cite{Amsler2012a} the $Cmcm$-disilane,~\cite{Flores-Livas2012} the new structure of LiAlH$_4$,~\cite{Amsler2012} the new polyneric phases of alanates,~\cite{Huan2013} etc, were discovered with the minima hopping method.  Novel structures of B,~\cite{Oganov2009} Na,~\cite{Ma2009} Ca,~\cite{Oganov2010} GeH$_4$~\cite{Gao2008}, NaH$_n$ (N$>$1),~\cite{Baettig2011} LiBeB,~\cite{Hermann2012} etc, were predicted using the EA. Based on the PSO algorithm, Zhu \textit{et al.} determined the long-puzzled high pressure structures of Bi$_2$Te$_3$.~\cite{Zhu2011} The new high-pressure structures of Li,~\cite{Lv2011} H$_2$O,~\cite{Wang2011} CaH$_6$,~\cite{Zhu2012} etc, were also successfully predicted using the PSO algorithm. High-pressure phases of silane SiH$_4$,~\cite{Pickard2006}  Al,~\cite{Pickard2010} and so on, were predicted by the random search technique.~\cite{Pickard2006,Pickard2011} All these cases proved the success of the CSP algorithms in the determination of crystal structures even when experimental data are unavailable. Nowadays, the theoretical predictions can guide the experimental syntheses. Meanwhile, the CSP technique can be a powerful tool of the theoretical design of materials.

The ideas of some optimization algorithms come from natural processes. For example, the EA was inspired by biological evolution: the process of whole population's evolution is the process of the survival of the fittest. The operators of EA mimic the mutation, reproduction, recombination, and selection of biological evolution.  For CSP, this algorithm is realized by introducing similar evolutionary operators. Similar to the theory of biological evolution, we often call a particular candidate structure as an individual, and the set of all structures generated in a single iteration of the algorithm as the generation.~\cite{Lonie2011} All generations form the crystal population.~\cite{Lonie2011} The modification of an individual is accomplished by the variation operators. The evolution direction and speed of the crystal population are up to the evolutionary operators, including the variation and the selection operators. The variation operators control the diversity of the crystal population, and the selection operator mainly determine the direction of evolution. The idea of simulated annealing (SA) algorithm was from the natural process of annealing in metallurgy, which hastens a melted material to be crystal by controlled cooling. The SA algorithm improves its efficiency through the introduction of two tricks: the so-called Metropolis selection rule~\cite{Metropolis1953} and lowering the ``temperature''. The basin hopping (BH) algorithm is just as its name that means trying to escape from the local minima by hopping.

All the optimization algorithms are stochastic and have their own advantages and disadvantages. The EA is powerful in global optimization and multi-objective optimization, but it is week in local optimization. While, the BH algorithm that is based on atoms' small random moves has better local optimization ability. It has no exchange of information between different individuals. The SA has the good global optimization ability and it can easily control the choice of high-energy structures to increase the population diversity. But it strongly depends on the initial structures. If we combine some of these algorithms to let them work collaboratively, we can manage to overcome disadvantages and make full use of their advantages. I refer to this idea as the multi-algorithm collaborative (MAC) crystal structure prediction. Presently, I combined three algorithms: the EA, SA, and BH algorithms, to construct a Multi-algorithm-collaborative Universal Structure-prediction Environment ({\sc Muse}). To further strengthen the search ability of {\sc Muse} in future, I keep it open and other algorithms can be included. The reasons why the three algorithms are chosen for combination are: firstly, some higher energy structures should be kept for crystal population diversity using the SA algorithm. Secondly, to prevent the solution from being trapped in local minima we can help it to escape from the local minima using the BH algorithm. Thirdly, the three algorithms are conceptually simple and can be combined easily.

In this paper, I focus on the implementation of MAC technique and how it works. This paper is organized as follows. Section~\ref{compdet} is the algorithm implementation. I will show some results predicted by {\sc Muse} in Section~\ref{results}. The predicted systems include metallic, covalent, and ionic systems. The conclusions are presented in Section~\ref{disconcl}.

\section{Algorithm implementation}
\label{compdet}
{\sc Muse} is written in Python and uses MAC algorithm to search the stable structures of materials. The MAC algorithm is expected to overcome the shortcomings of an individual algorithm and improve the search efficiency. Furthermore, the evolution of the crystal population and the choice of the operators are designed to be self-adaptive (For details, see Section~\ref{sae}). In other words, in the search, the crystal population will undergo a self-adaptive evolution process. So, {\sc Muse} seeks materials' stable structures effectively under certain conditions.

The implementation of MAC crystal structure prediction is based on the manipulation of crystal structures in real space, i.e., the operations on the crystal structures made by the variation operators. A crystal structure is described by the shape of the lattice cell and the atoms positions therein. In the evolution process of crystal population, the fitness function can be free energy, hardness, volume, band gap, and so on. At present, {\sc Muse} can predict the stable structures according to energy criterion. The fitness function is thus the free energy of each structure. The structure with lowest free energy has the best fitness. The selection operator is to select the low free energy structures for variation operators to generate new structures.

{\sc Muse} depends on external local optimization codes. At present, {\sc Muse} uses the Vienna Ab initio Simulation Package (VASP),~\cite{Kresse94,Kresse96} SIESTA,~\cite{Soler2002} Quantum ESPRESSO,~\cite{pwscfcite} and LAMMPS~\cite{lammps} as the local optimization tools. It determines the space group number of each optimized structure immediately after the optimization is done. The duplicate structures are eliminated according to the space group numbers and the nearest triangles formed by the nearest three atoms (For details, see Section~\ref{ed}). More importantly, {\sc Muse} generates the random structures of the first generation with symmetry constraints, which largely shortens the local optimization time of the first generation and increases the diversity of crystal structures. The random structures are created according to the randomly chosen space group numbers from 2 to 230, and Wyckoff positions must be fit to the atom numbers ratio. Especially for large systems, the constraints on symmetry will avoid unphysical disorder crystal structures, similar to glass state.  {\sc Muse} can also pick up (restart) a previous interrupted search from where it stopped.

\begin{figure}[htp]
\centering
\includegraphics[scale=0.50,trim=1cm 1.5cm 1.5cm 0cm]{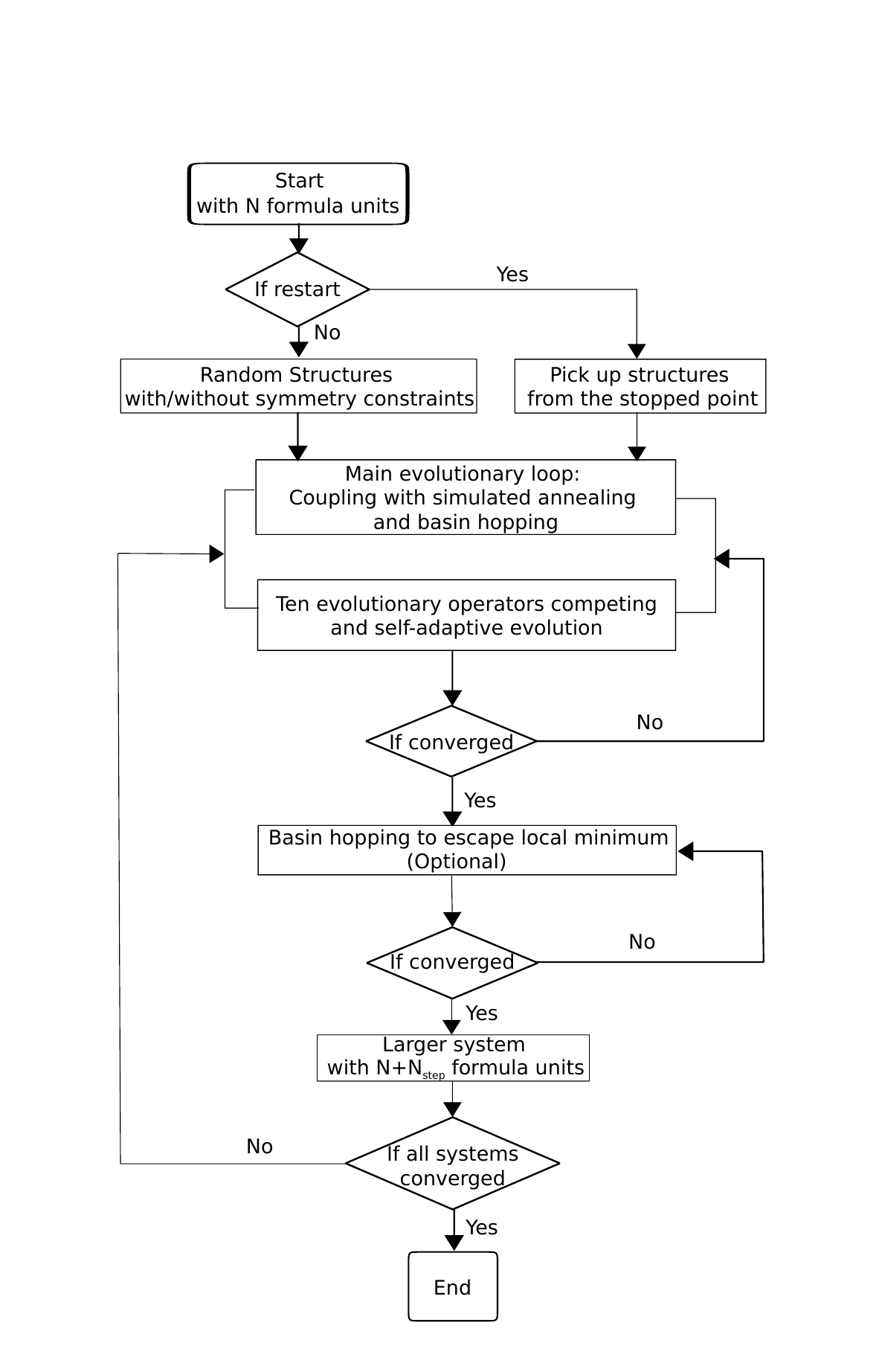}
\caption{The flowchart of multi algorithms collaboration and coupling in {\sc Muse}. N$_{\mathrm{step}}$ means the increase number of formula units to enlarge the supercell.}
\label{fig:flowchart}
\end{figure}

\subsection{Multi-algorithm collaboration}
A single algorithm has more or less inherent disadvantages. Using the combination of multi algorithms, or hybrid, we can overcome their disadvantages and make full use of their advantages. The efficiency of the MAC search also depends on the diversity of crystal structures. To increase the diversity of crystal structures, I created ten variation operators in {\sc Muse}: cross over, mutation, permutation, cross-over-mutation, permutation-mutation, slip, twist, random move, ripple, and mutation-ripple. The multi-algorithm collaboration and coupling are shown in Fig.~\ref{fig:flowchart}.

Whether the optimized structure is kept or not in the main loop is controlled by the Metropolis algorithm.~\cite{Metropolis1953} This couples the EA and SA algorithms. To avoid prematuration, I coupled the main loop with BH algorithm (Fig.~\ref{fig:flowchart}). After the main loop is converged, {\sc Muse} can turn into the pure BH loop to help the solution to escape from local minima if it is possibly trapped in local minima in the main loop. While, the pure BH loop is optional and we can run the main loop without the pure BH loop. As we know, the smaller systems that have smaller number of formula units often converge faster than larger ones, because they have smaller configurational space. For larger systems, {\sc Muse} has the trick to prepare seed structures from the converged smaller systems. To test this trick, I used the 12-atom LiBC system. The number of structures in each generation is 20. Tests show if we pick out the low-energy structures from the smaller system (6 atoms) to prepare seeds, we can easily converge the larger system (12 atoms) with the help of seeds (See Table~\ref{testseeds}). So we are encouraged to go from smaller systems to larger ones. This process is automatic in {\sc Muse}. If all the defined systems are converged, {\sc Muse} will terminate the search and exit.

\begin{table}
\caption{Tests of seeds on search efficiency. The test system is LiBC at 0 GPa. N$_{\mathrm{atom}}$ is the number of atoms in the system. P$_{\mathrm{size}}$ is the population size. N$_\mathrm{run}$ is the number of runs for statistics. Mean N$_{\mathrm{best}}$ is the averaged generation number to find the lowest-energy structure. The seeds were prepared from the 6-atom search.}
\label{testseeds}
\begin{tabular}{ccccccc}
\hline\hline
If use seeds & N$_\mathrm{atom}$ & P$_{\mathrm{size}}$ &N$_\mathrm{run}$ & Mean N$_{\mathrm{best}}$& Success \%\\
\hline
	No & 12  & 20 & 3 & 4.50 & 100\\
	Yes & 12 & 20 & 3 & 1.00 & 100\\
\hline\hline
\end{tabular}
\end{table}

\subsection{The first generation}

A unit cell in {\sc Muse} is described with six parameters, $\mathbf a$, $\mathbf b$, $\mathbf c$, $\alpha$, $\beta$, and $\gamma$, and the atoms positions, where $\mathbf a$, $\mathbf b$, and $\mathbf c$ are the lattice vectors and $\alpha$, $\beta$, and $\gamma$ are the corresponding angles. The structures of the first generation can be randomly created with the symmetry constraints  and the constraints of the minimum and maximum angles (45$^\circ$ and 135$^\circ$) between lattice vectors. For unbiased search, the space group number of the generated structure can be randomly chosen from 2 to 230 with equal possibility. Atoms are placed in the corresponding positions according to symmetry operations with Wyckoff positions constraint. If we want to generate the first generation in the fully random manner, {\sc Muse} will generate the structures without symmetry constraints. Then its volume is scaled to the coarsely guessed value specified in the input file. If a space group number has been used, {\sc Muse} will not choose it for the rest of the random structures to avoid duplication and thus increase the diversity. 

\subsection{Self-adaptive evolution}
\label{sae}
In the evolution process of crystal structures, different systems attempt to seek their optimal variation operators to achieve the possible largest diversities. This scheme is expected to increase the search efficiency and success rate. So from the second generation, the ten variation operators will compete in their success rates. Each operator has the initial success rate of 100\% . The operators are chosen to generate new structures according to their success rates. The more positive contributions to the population an operator has, the larger success rate it gets. The positive contribution means that it produced lower enthalpy structure or changed the symmetry/symmetries of parent structure/structures. So, the success rate of an operator is calculated via $\frac{N_{\mathrm{success}}}{N_{\mathrm{called}}}$, where $N_{\mathrm{success}}$ and $N_{\mathrm{called}}$ are the number of its success and the number that it has been called in the last generation, respectively. The larger success rate an operator has, the larger possibility it has to breed offspring. This is called the self-adaptation of variation operators. That is to say, the survival of the fittest is also applied to the selection of operators. This type of self-adaptive evolution turns out to increase the diversity of structures and hasten the convergence of global search (See Section~\ref{results}).
 
\subsection{Variation operators}
The individuals after the first generation are created from the last generation using ten variation operators. The ten operators fall into two categories: the single-parent based operators and the two-parent based ones. The mutation, permutation, random move, ripple, slip, twist, permutation-mutation, and mutation-ripple operators are all the single-parent based operators. The cross over and cross-over-mutation operators are two-parent based operators.

\subsubsection{Mutation}

\begin{figure}[htp]
\centering
\includegraphics[scale=0.20]{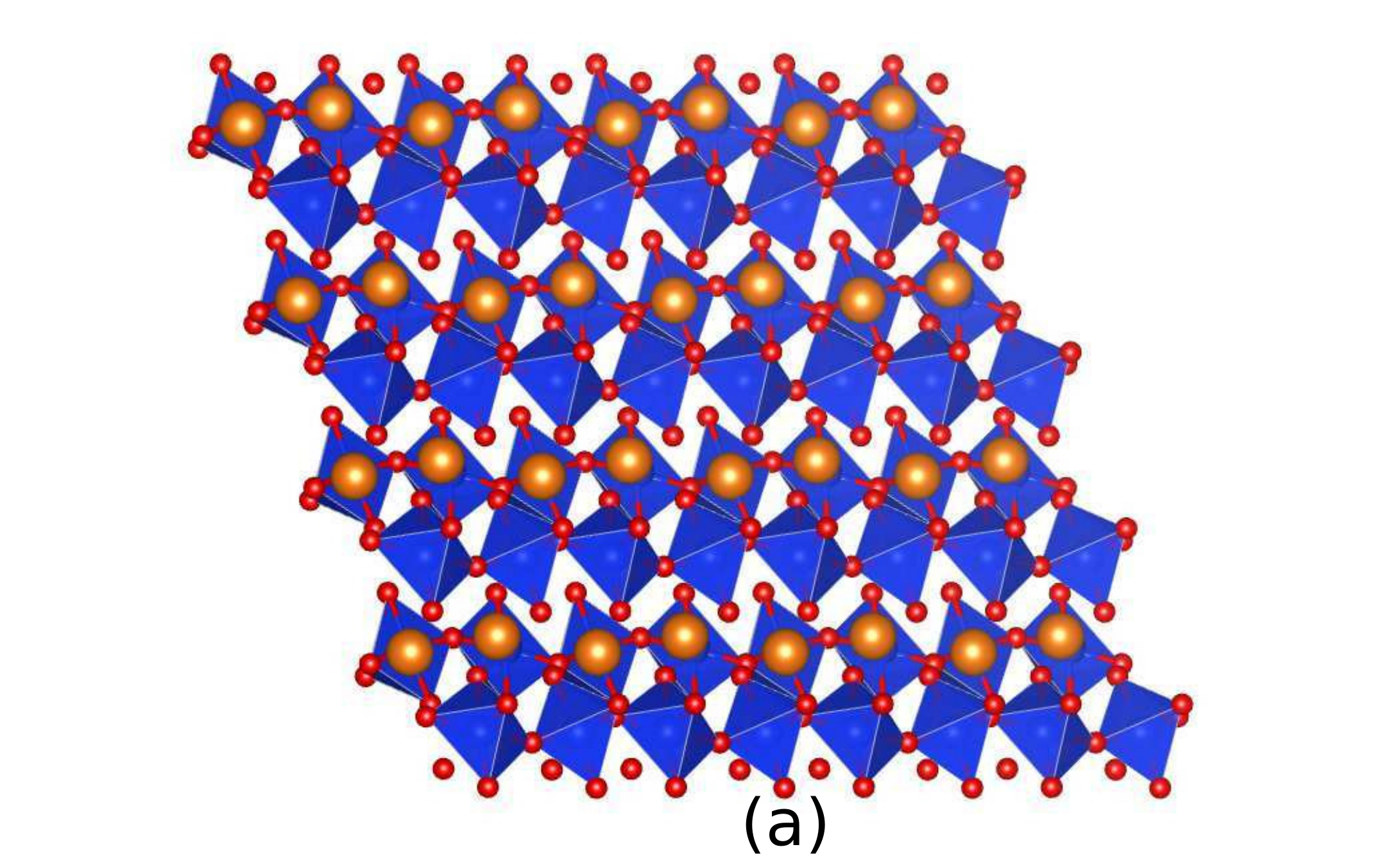}
\includegraphics[scale=0.37,trim=0cm 1.5cm 3.2cm 0cm]{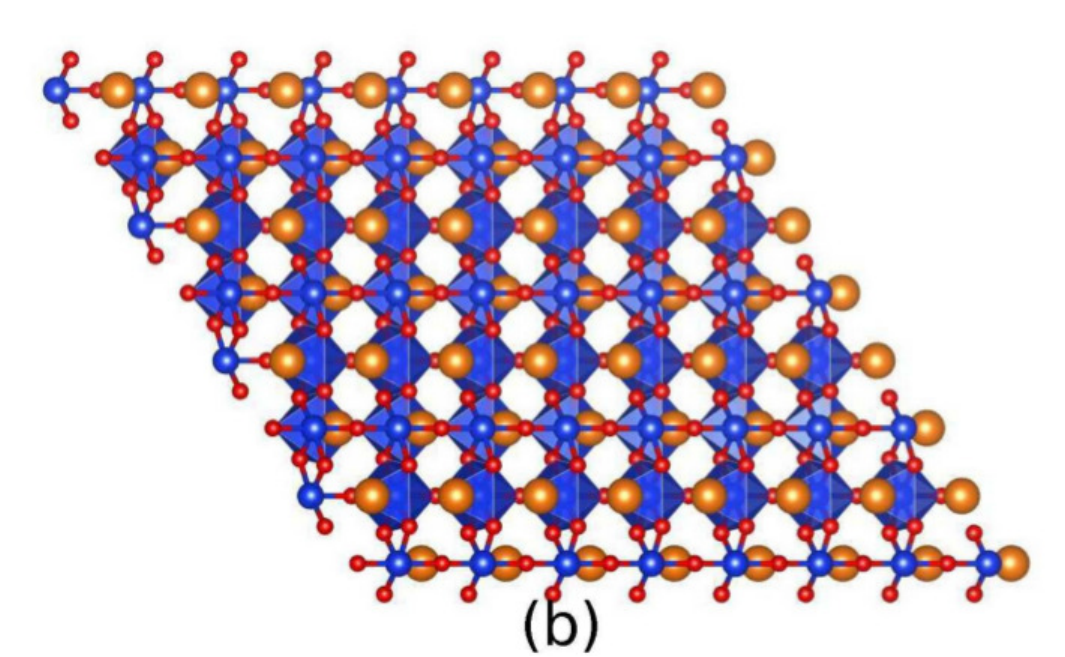}
\caption{The example of mutation. (a) Randomly chosen parent: $P-1$ (2), (b) Offspring: $P2_1/m$ (11). The two structures are locally optimized.}
\label{fig:mutation}
\end{figure}

The mutation operator variates one selected parent to produce a new individual by multiplying the unit cell row vector with a symmetric Voigt strain matrix:~\cite{Oganov2006-2,Oganov2006-3,Lonie2011}
\begin{equation}
\mathbf{a'}=\left[ \begin{array}{ccc}
	1+\delta_{11} & \delta_{12}/2 & \delta_{13}/2\\
	\delta_{12}/2 & 1+\delta_{22} & \delta_{23}/2\\
	\delta_{13}/2 & \delta_{23}/2 & 1+\delta_{33}\\
\end{array} \right] \mathbf{a},
\end{equation}
where $\delta_{ij}$s are the zero-centered Gaussian random variables with a specified standard deviation. $\mathbf a$ is the original lattice vector, and $\mathbf a'$ is the new lattice vector. The atomic coordinates are fractional values and are then scaled to the new vectors. Actually, the mutation operator is to apply a shear strain on the cell to cause it to undergo phase transition. So it is a very effective operation to diversify the crystal population. Because different systems have different optimal standard deviations $\delta_{ij}$, {\sc Muse} can use trial and error scheme to obtain good diversity from the initial value. The standard deviation value can also be fixed. After applying the strains, the crystal's volume is then scaled to the volume of the lowest enthalpy structure in the previous generation. Figure~\ref{fig:mutation} shows an example of mutation.

\subsubsection{Permutation}
The permutation operator changes the positions of two kinds of atoms a random number of times.~\cite{Oganov2006-2,Oganov2006-3,Lonie2011} This obviously increases the diversity of the crystal population after operation. Only when the atom types are more than one can this operator be used.

\subsubsection{Permutation-mutation}
The permutation-mutation operator is the hybrid of the permutation and mutation operators. The unit cell is mutated after permutation operation. This operator is a single-parent one.

\subsubsection{Random move}
The random move operator moves all atoms positions randomly, keeping the lattice vectors unchanged. It displaces the fractional coordinates of all atoms by random amount between [-$D_{\mathrm{max}}$, $D_{\mathrm{max}}$], where $D_{\mathrm{max}}$ is the maximum percentage of coordinates that atoms displaced. I realized the BH algorithm in {\sc Muse} by combining this operator with the Metropolis rule.~\cite{Metropolis1953,Wales1999} The periodic boundary conditions are applied to the unit cell after operation.

\subsubsection{Slip}
\begin{figure}[htp]
\centering
\includegraphics[scale=0.40]{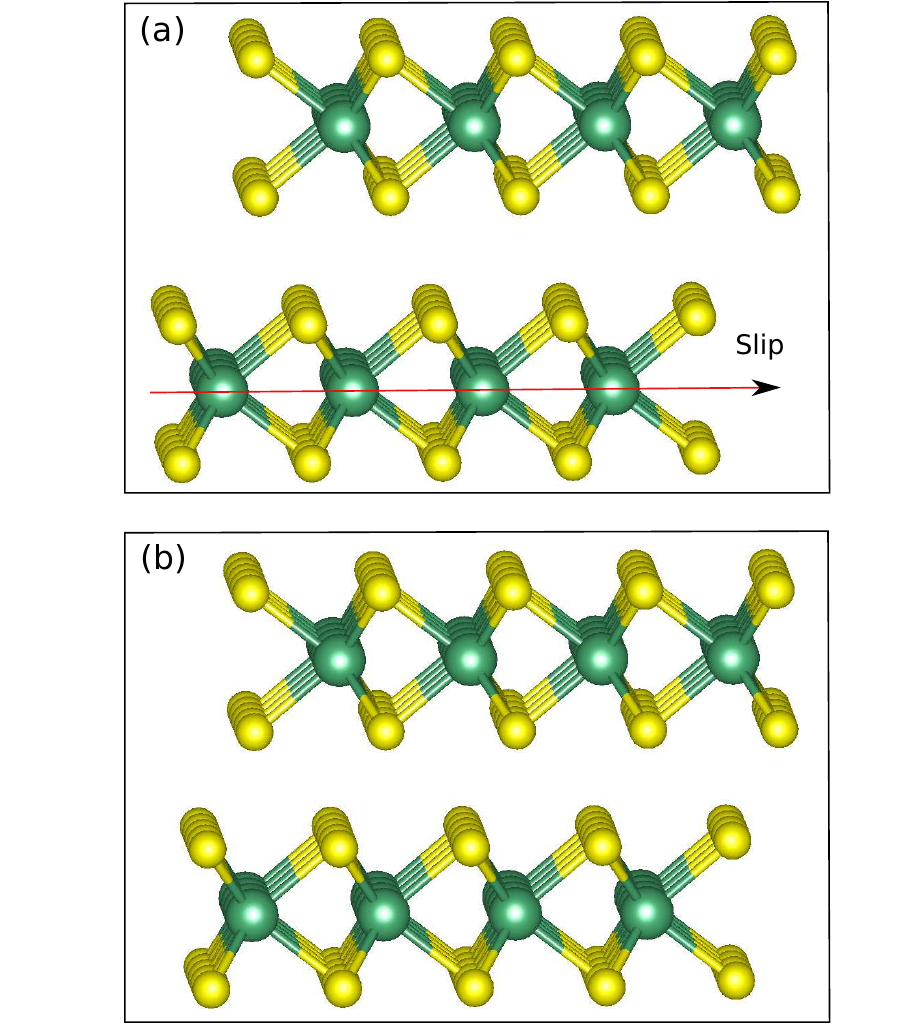}
\caption{The slip operator. The new structure is obtained by slipping several layers of atoms of the older structure. (a) before slip. (b) after slip.}
\label{fig:slip}
\end{figure}

To fully diversify crystal structures, I introduced a new operator, slip, into {\sc Muse}. This operator slips a group of atoms by a random distance along a random direction that is parallel to a specific plane (Fig.~\ref{fig:slip}). This group of atoms are chosen when their fractional coordinates of the randomly selected axis are greater than 0.5. The idea of slip operator comes from the real crystal phase transition, in which some atom layers slip along special directions. The structures of some crystals transit by slipping specific atom layers, such as ZnS~\cite{Akizuki1981} and MoSi$_2$.~\cite{Mitchell1999} So the slip operator is expected to increase the diversity of crystal structures.

\subsubsection{Twist}
\begin{figure}[htp]
\centering
\includegraphics[scale=0.28]{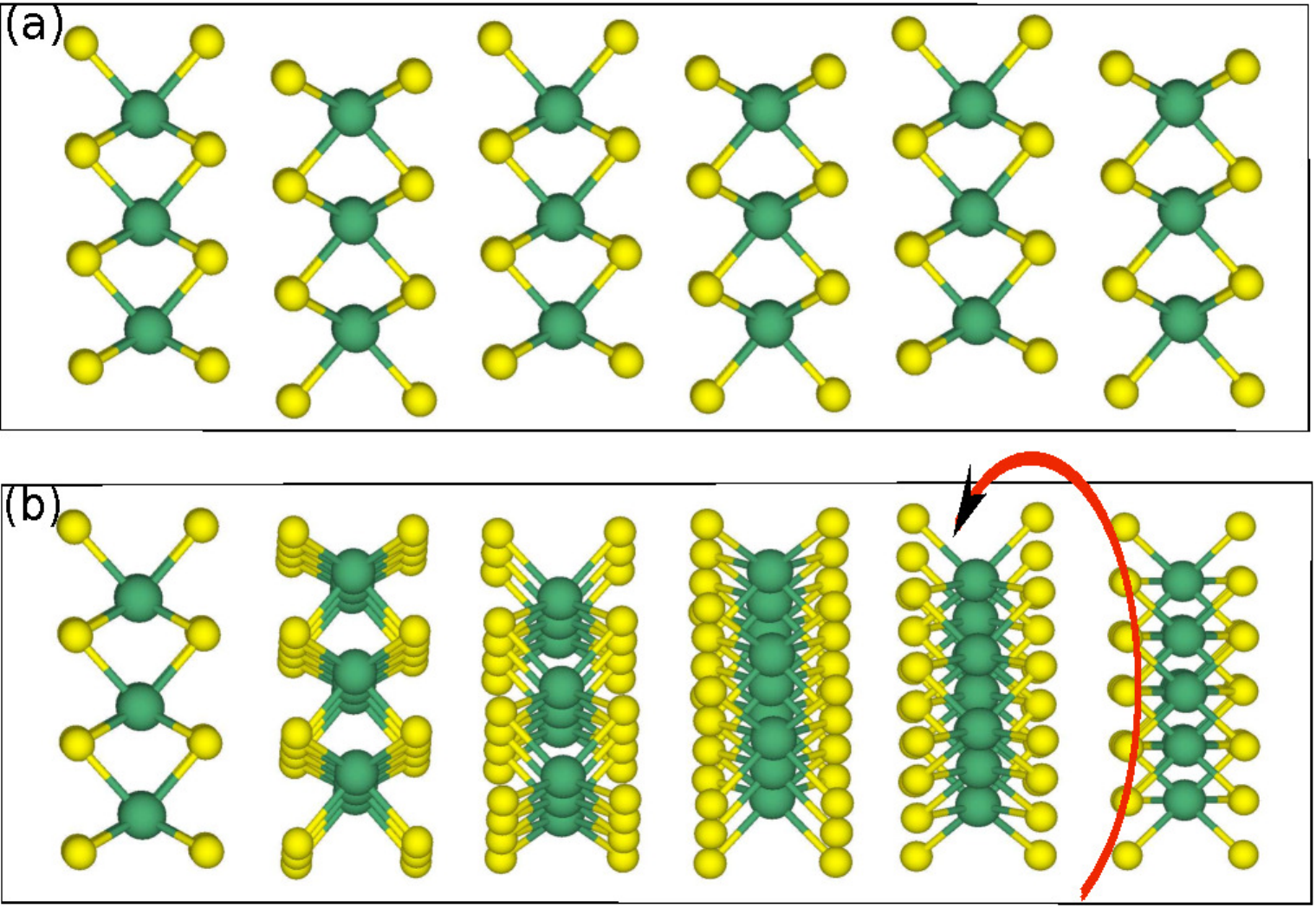}
\caption{The twist operator. (a) before twist. (b) after twist.}
\label{fig:twist}
\end{figure}

The twist operator is also a new one. It twists a crystal cell by rotating atoms around a randomly chosen axis. The rotation angles of atoms increase with the increasing fractional coordinates of the chosen axis. If the first atoms rotated by $\Delta \theta$, the second atom will rotate by $2\Delta \theta$, and the third $3\Delta \theta$, etc. $\Delta \theta$ is $\pi/8$. During the rotation, the cell is fixed and only the fractional coordinates of the atoms are changed. Figure~\ref{fig:twist} shows the general idea of the twist operator. In order to show the twisted atoms clearly, I rotated several layers of atoms together. For smaller primitive cells, the twist operator is expected to slightly increase the diversity of the crystal structures. However it greatly enhances the diversity of crystal structures for larger supercells because of larger configurational space. Twist also causes phase transition in real materials, such as molecular crystals.~\cite{Saito1995,Knowles2012} The twisted bilayer graphenes are formed by twisting two adjacent layers.~\cite{Moon2013}

\subsubsection{Ripple}
The ripple operator, first proposed by Lonie and Zurek,~\cite{Lonie2011} is the periodic displacement operator which shifts the coordinates of each atom along a randomly chosen axis by certain amount. For example, if $x$-axis is chosen, the $x$ components of all atoms will be shifted to the new values in fraction: $x_{\mathrm {new}} = x + \Delta x$, where $\Delta x$ depends on the atom’s non-displaced (i.e., $y$ and $z$) coordinates via:~\cite{Lonie2011}
\begin{equation}
\Delta x = \rho \mathrm{cos}(2\pi \mu y+\theta_y) \mathrm{cos}(2\pi \eta z+\theta_z),
\end{equation}
where $\rho$ is the maximum possible displacement in $x$ direction, and $\mu$ and $\eta$ are integers, and $\theta_y$ and $\theta_z$ are random numbers between [0, 2$\pi$).

\begin{figure}[h]
\centering
\includegraphics[scale=0.18]{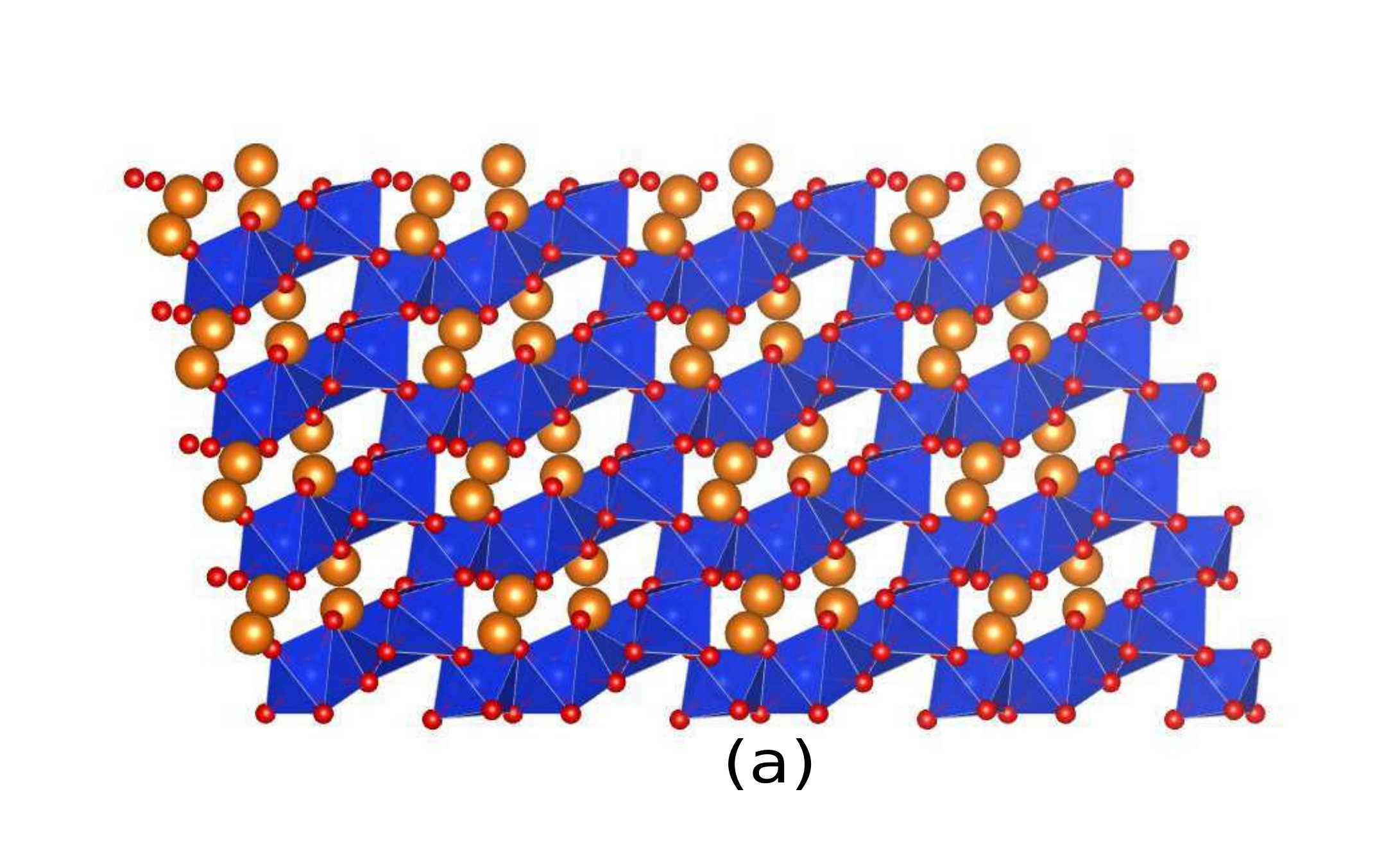}
\includegraphics[scale=0.16]{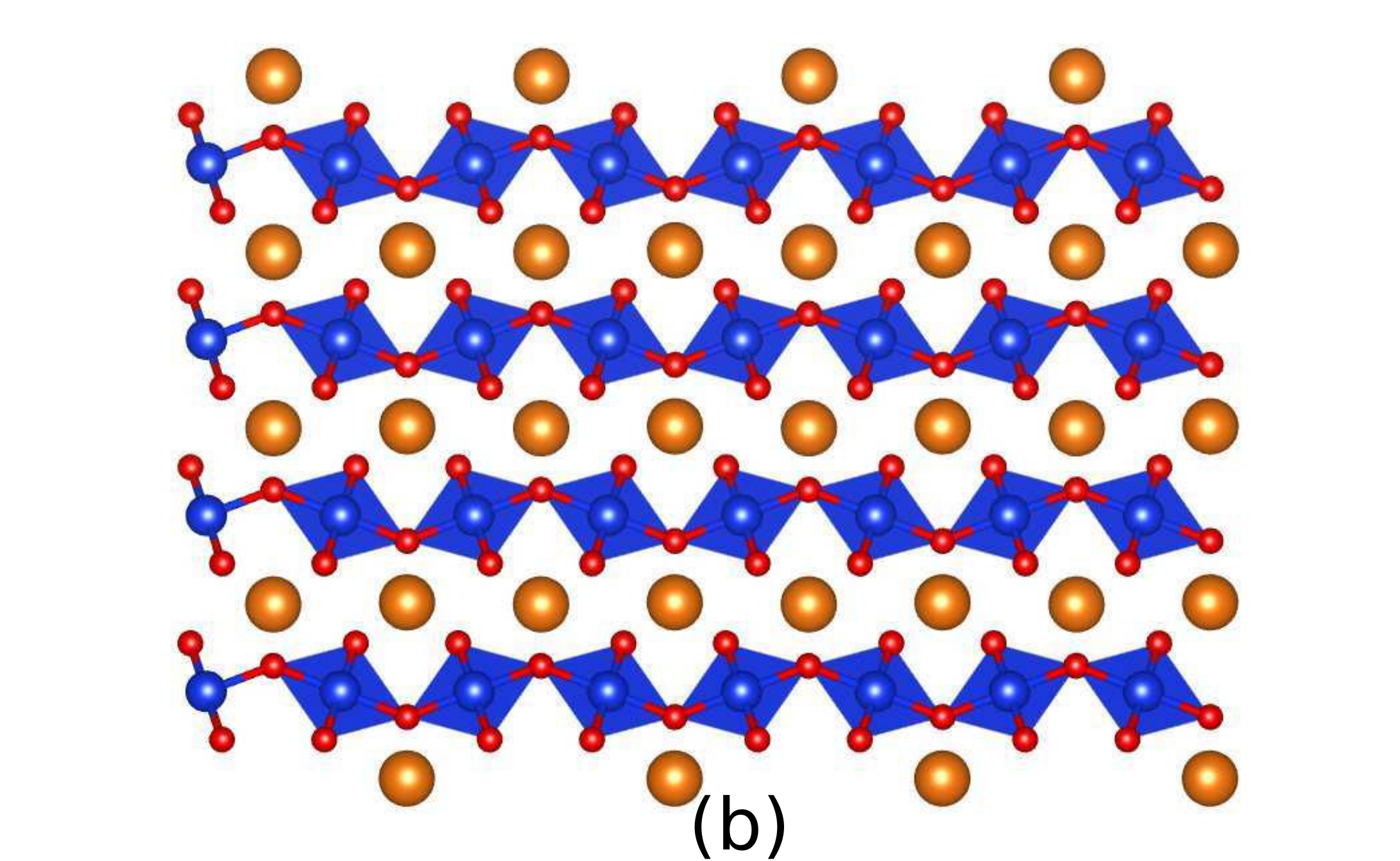}
\caption{The example of ripple. (a) Randomly chosen parent: $P1$ (1), (b) Offspring: $Cmcm$ (63). The two structures are locally optimized.}
\label{fig:ripple}
\end{figure}

The ripple operator is also a singe-parent operator, but it can increase the crystal diversity substantially.~\cite{Lonie2011} Some solids manifest the ripple motif at ambient or high pressure, such as Cs-III,~\cite{McMahon2001} Rb-III,~\cite{Nelmes2002} and Ga-II.~\cite{Degtyareva2004} Figure~\ref{fig:ripple} is an example of ripple.

\subsubsection{Mutation-ripple}
This operator is the hybrid of the mutation and ripple operators. The unit cell is manipulated by the ripple operator after it is mutated.

\subsubsection{Cross over}
\begin{figure}[htp]
\centering
\includegraphics[scale=0.23]{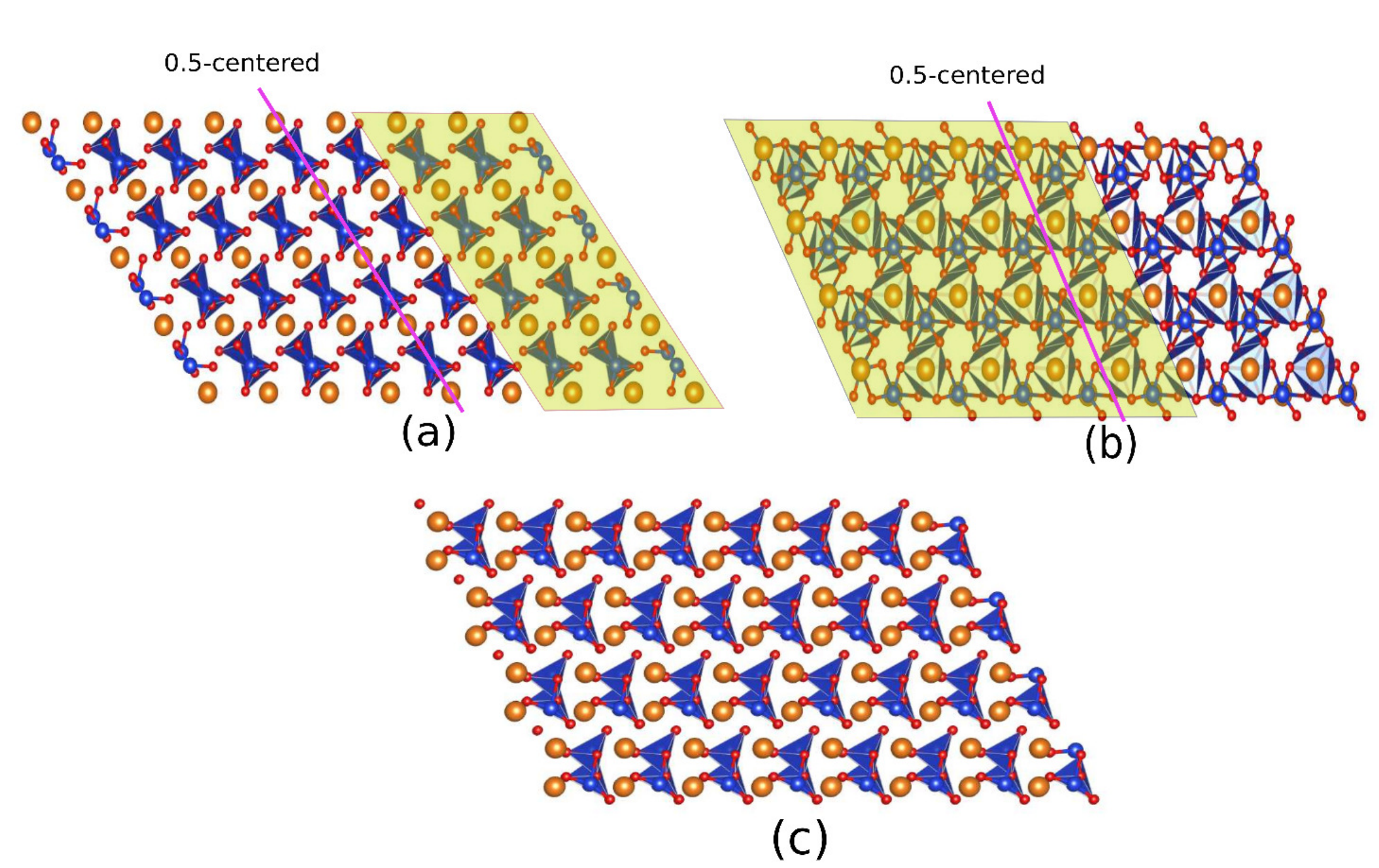}
\caption{The example of cross over. (a) Randomly chosen parent 1: $Ccmm$ (63), (b) Randomly chosen parent 2: $P6_3/m$ (176), (c) Offspring: $Pm$ (6). The three structures are all locally optimized.}
\label{fig:cross}
\end{figure}

The cross over operator is a two-parent based operator. It cuts and splices the randomly picked two parents to make a new individual.~\cite{Deaven1995,Oganov2006-2,Oganov2006-3,Woodley2007,Assadollahzadeh2008,Abraham2006,Trimarchi2007,Hooper2009,Lonie2011}
It mainly consists of two steps: cut and splice. In the first step, the cut plane is parallel to the randomly not picked two vectors (e.g., $\mathbf{b}$ and $\mathbf{c}$) of one parent. The cut position is in the randomly picked vector (i.e., $\mathbf{a}$) and the cut fractional coordinate is the 0.5-centered Gaussian random number between [0.25, 0.75]. If the atoms coordinates of the picked vector (i.e., $\mathbf{a}$) is smaller than the Gaussian random number, they are taken as the part of the new individual (the left part of Fig.~\ref{fig:cross} a). The other part of the new individual is from the other parent in the same way but the atoms coordinates are larger than the same Gaussian random number (the right part of Fig.~\ref{fig:cross} b). In the second step, the picked atoms from the two parents are placed together to make the new individual (Fig.~\ref{fig:cross} c). Its lattice vectors are the vector summation of the lattice vectors of two parents. To increase the success rate, the atoms are centered before the cross over operation.

\subsubsection{Cross-over-mutation}
The cross-over-mutation operator is the hybrid of the cross over and mutation operator.

\subsection{Evaluation and selection}
As other stochastic structure search algorithms for stable structures,~\cite{Oganov2006-3,Lonie2011,Oganov2010,Wang2010,Wang2012,Lv2012,Pickard2011}  {\sc Muse} also uses \textit{ab initio} Gibbs free energy as the fitness function, which evaluates if an individual is
suited to be parent to breed offspring. At 0 K, Gibbs free energy equals to enthalpy. The enthalpy ($H=E+PV$) of an individual is determined from the local optimization code. The absolute probability $p_i$ of an individual being selected for breeding offspring is
determined by
\begin{equation}
p_i = \frac{H_\mathrm{max}-H_{\mathrm{i}}}{H_{\mathrm{max}}-H_{\mathrm{min}}},
\end{equation}
where $H_i$ is the enthalpy of this individual, and $H_{\mathrm{max}}$ and $H_{\mathrm{min}}$ are the enthalpies of the worst and the best individuals in the last generation, respectively.

After an offspring is generated and locally optimized, whether it is kept or not is controlled by the Metropolis rule:~\cite{Metropolis1953}

$\bullet$ An individual is kept if its enthalpy is lower than that/those of its parent(s).

$\bullet$ But if its enthalpy is greater than that/those of its parent(s), it is kept only if 
\begin{equation}
\mathrm{exp}[{(H_{\mathrm{parents}} - H_{\mathrm{individual}})/kT}]
\end{equation}
 is greater than a randomly picked number from [0, 1]. $k$ is Boltzmann constant. The initial temperature $T_0$ for annealing is specified in the input file. The temperature in generation $\mathrm{n+1}$ is reduced to $T_\mathrm{n+1} = 0.9T_\mathrm{n}$.

\subsection{Elimination of duplicates}
\label{ed}
\begin{figure}[htp]
\centering
\includegraphics[scale=0.15]{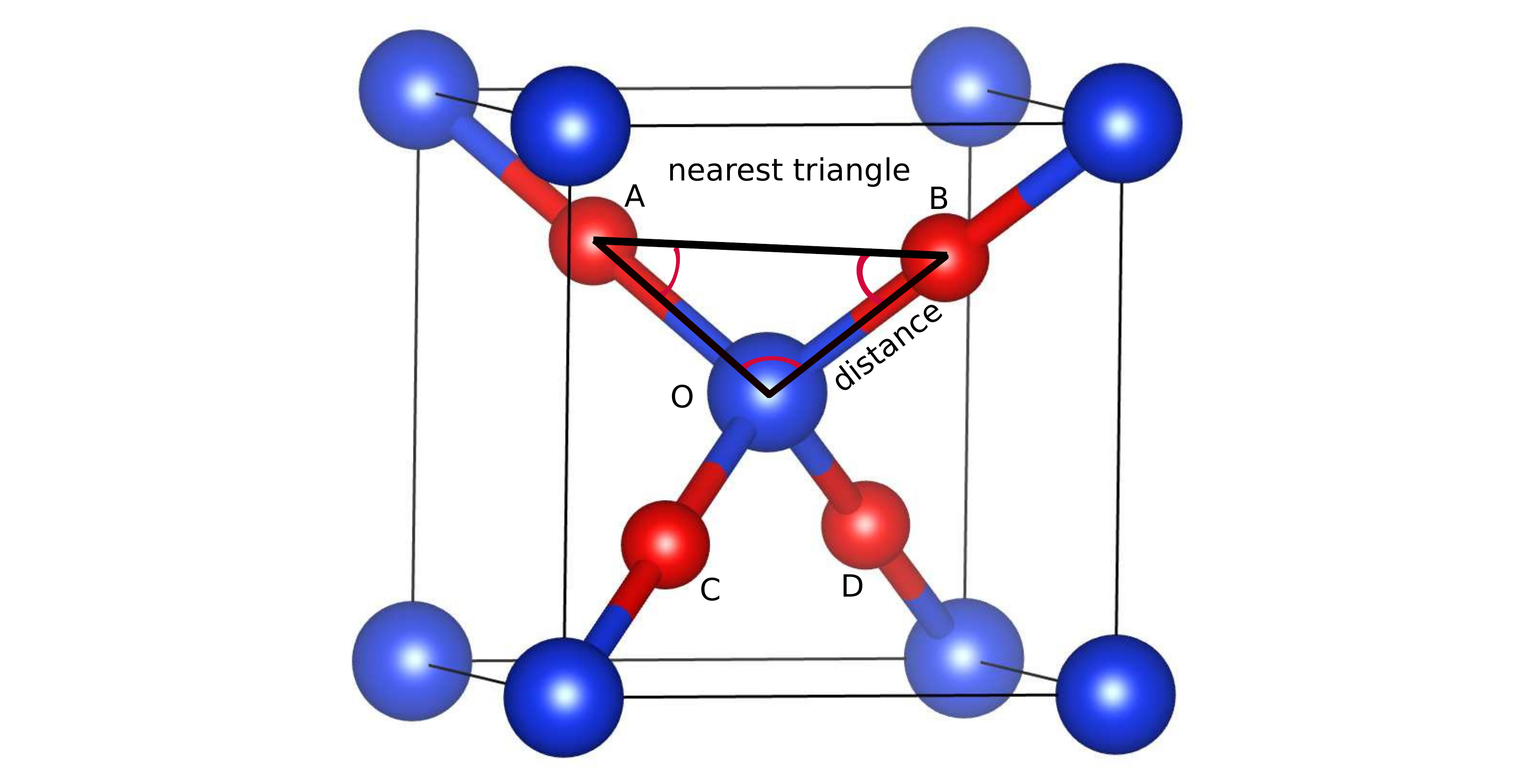}
\caption{The nearest triangles formed by the reference atom and its first/second nearest neighbors.}
\label{fig:nearest}
\end{figure}

The duplicate structures prohibit the diversity of crystal population and then decrease the search efficiency and success rate. In {\sc Muse}, I developed a new method to delete duplicates. This method is based on the determination of space group numbers and nearest triangles formed by the nearest atoms. In detail, in the first step {\sc Muse} judges if two structures have the same space group number within the specified tolerance of distance, usually 0.1 $\mathrm{\AA}$. If they do not have, they are treated as different structures. Otherwise, the two structures are initially reduced to primitive cells. Then {\sc Muse} judges whether the numbers of atoms in the two primitive cells are equal. If not equal, they are different structures. If equal, it will further compare the nearest distances and the coordination numbers of every inequivalent atom. The inequivalent atoms are determined according to symmetry operations. If the nearest distances and the coordination numbers of the two primitive cells are equal for every inequivalent atom correspondingly, {\sc Muse} compares the nearest triangles formed by three atoms, i.e., the reference atom and the first/second nearest neighbors (see Fig.~\ref{fig:nearest}). If all the triangles of every inequivalent atom in the two primitive cells are congruent correspondingly within the specified tolerances of distance and angle, they are treated as the same structure. Then the duplicated one with higher enthalpy is then deleted. If the shape formed by the three atoms is not a triangle but a line, we can note its angles as 0$^\circ$, 0$^\circ$, 180$^\circ$. 

\subsection{Termination}
The termination of the main evolutionary loop is controlled by the terminator operator. The main loop is terminated:

$\bullet$ if the number of continuous generations with the same symmetry structures whose best enthalpy differences are less than 1 meV/atom reaches the number specified in the input file;

$\bullet$ if the specified maximum number of generations has been done;

$\bullet$ if the diversity of the crystal population is too low.

The pure BH loop is optional. When it is used, it is terminated when the lowest-enthalpy structure and its enthalpy do not change any more in the specified number of generations. It helps the search to escape from local minima by moving atoms randomly . But usually we can only run the main loop without the pure BH loop.

\subsection{K-point adaptation}
The K-point adaptation is similar to Ref.~\onlinecite{Oganov2006-3}. The K-grid in reciprocal space of a lattice vector $i$ is calculated via:
\begin{equation}
k_\mathrm{i} = \frac{1}{a_\mathrm{i} \cdot k_{\mathrm{resol}}},
\end{equation}
where $a_\mathrm{i}$ is the length of the lattice vector $i$, and $k_{\mathrm{resol}}$ is the reciprocal-space resolution set in the input file. $k_\mathrm{i}$ is then rounded to an integer.

\begin{table}
\caption{The stable structures of different systems predicted by {\sc Muse} under different pressures, $P$ (in GPa). Some parameters are the same as those in Table~\ref{testseeds}. N$_{\mathrm{EA}}$ is the number of structures generated by the evolutionary algorithm. N$_{\mathrm{opt}}$ is the total number of optimized structures in each case. N$_{\mathrm{opt}}-\mathrm{N}_{\mathrm{EA}}$ is the number of structures generated by the basin hopping algorithm. The standard deviation in the mutation operator is 0.5.}
\label{diffstr}
\begin{tabular}{rcccccrc}
\hline\hline
	System & N$_\mathrm{atom}$ & $P$ & P$_{\mathrm{size}}$ &N$_\mathrm{run}$ & N$_{\mathrm{EA}}$/N$_{\mathrm{opt}}$ & Symmetry & Mean N$_\mathrm{G}$\\
\hline
	GaN & 20 & 0 &30 & 3& 183/193& $P6_3mc$\cite{Yeh1992} & 2.14\\
	TiO$_2$ & 12 & 0 & 20 &3 & 223/243 & $I4_1/amd$~\cite{Burdett1987} & 4.05\\
	ReO$_3$ & 16 & 0 & 20 &3 & 186/192 & $Pm\bar3m$~\cite{Jorgensen1986} & 2.13\\
	Al$_2$O$_3$ & 10 & 0 & 20 &20 & 712/788& $R\bar3c$~\cite{Pauling1925} & 1.97\\
	LiBC & 12 & 0 & 20 & 3& 241/270 & $P6_3/mmc$~\cite{Fogg2006} & 4.50\\
	MgSiO$_3$ & 20 & 130 & 30 &3 & 550/632 & $Cmcm$~\cite{Ono2006} & 7.02\\
\hline\hline
\end{tabular}
\end{table}

\section{Results}
\label{results}
The efficiency of MAC search in {\sc Muse} bas been fully tested for many cases including metallic, covalent, and ionic systems. Table \ref{diffstr} shows the searched stable structures of some systems under different conditions. The statistical data were obtained by counting not less than three searches. The random structures in the first generation were constructed with symmetry constraints. The systems converged in the first generation only through symmetry constraints are not listed. The \textit{ab initio} optimizations and the free energy calculations for every structure generated by {\sc Muse} were performed with VASP.~\cite{Kresse94,Kresse96} The generalized gradient approximation (GGA) parametrized by the Perdew-Burke-Ernzerhof (PBE)~\cite{Perdew1996} was applied and the electron-ion interactions was described by the projector augmented wave (PAW) scheme.~\cite{Blochl94,Kresse99} The kinetic energy cutoff and the \textit{k}-point grids spacing were chosen to be 1.3 times the default values and 0.02 $\mathrm{\AA}^{-1}$, respectively. As one can see, the systems with not more than two kinds of atoms are easily converged. {\sc Muse} generally finds their stable structures in less than four generations. However, the systems with more than two kinds of atoms need at least four generations. All the searched stable structures of these systems are in agreement with their known structures.

\begin{figure}[htp]
\centering
\includegraphics[scale=0.19,trim=2cm 0cm 1cm 0cm]{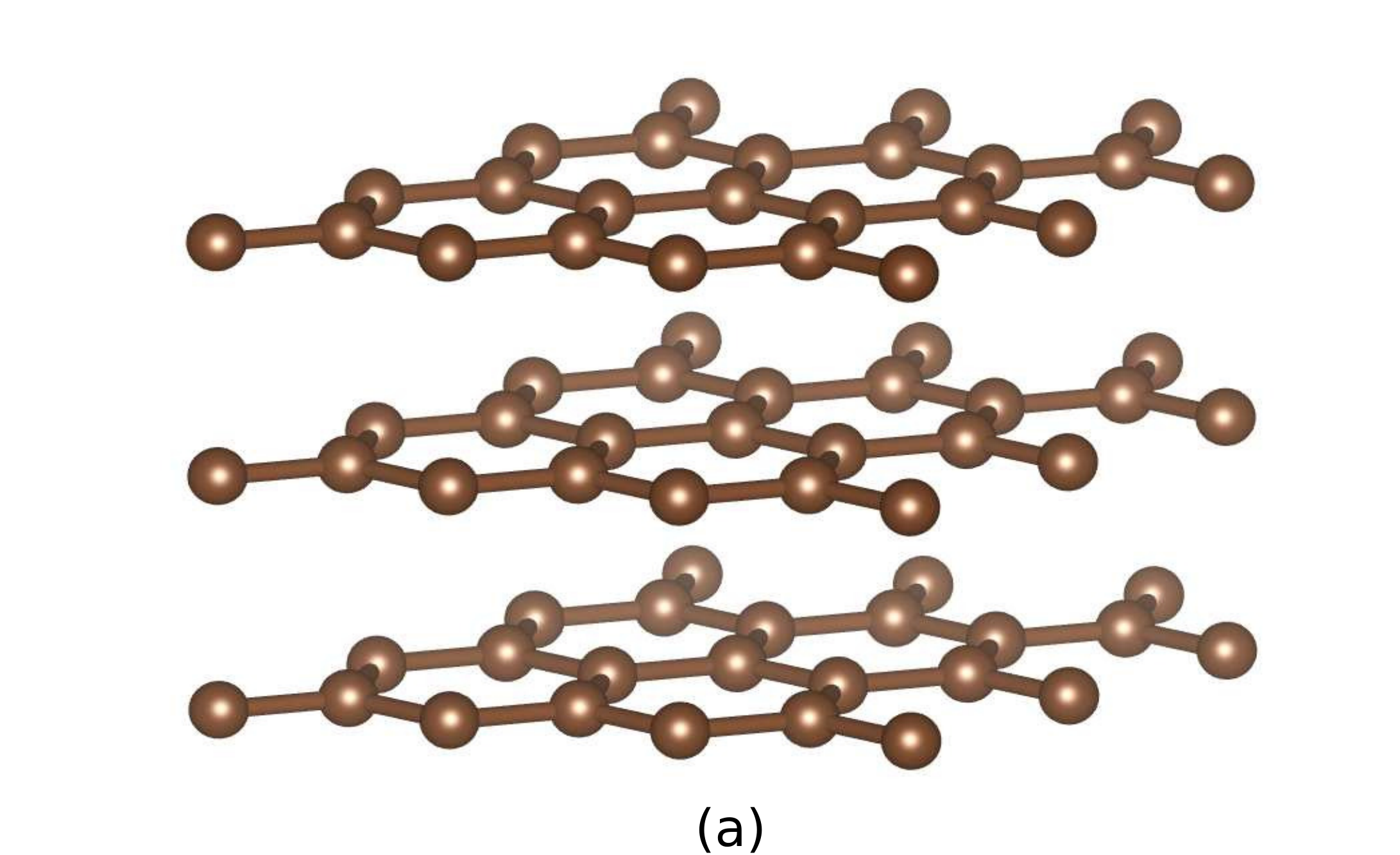}
\includegraphics[scale=0.18,trim=0cm 0cm 3cm 0cm]{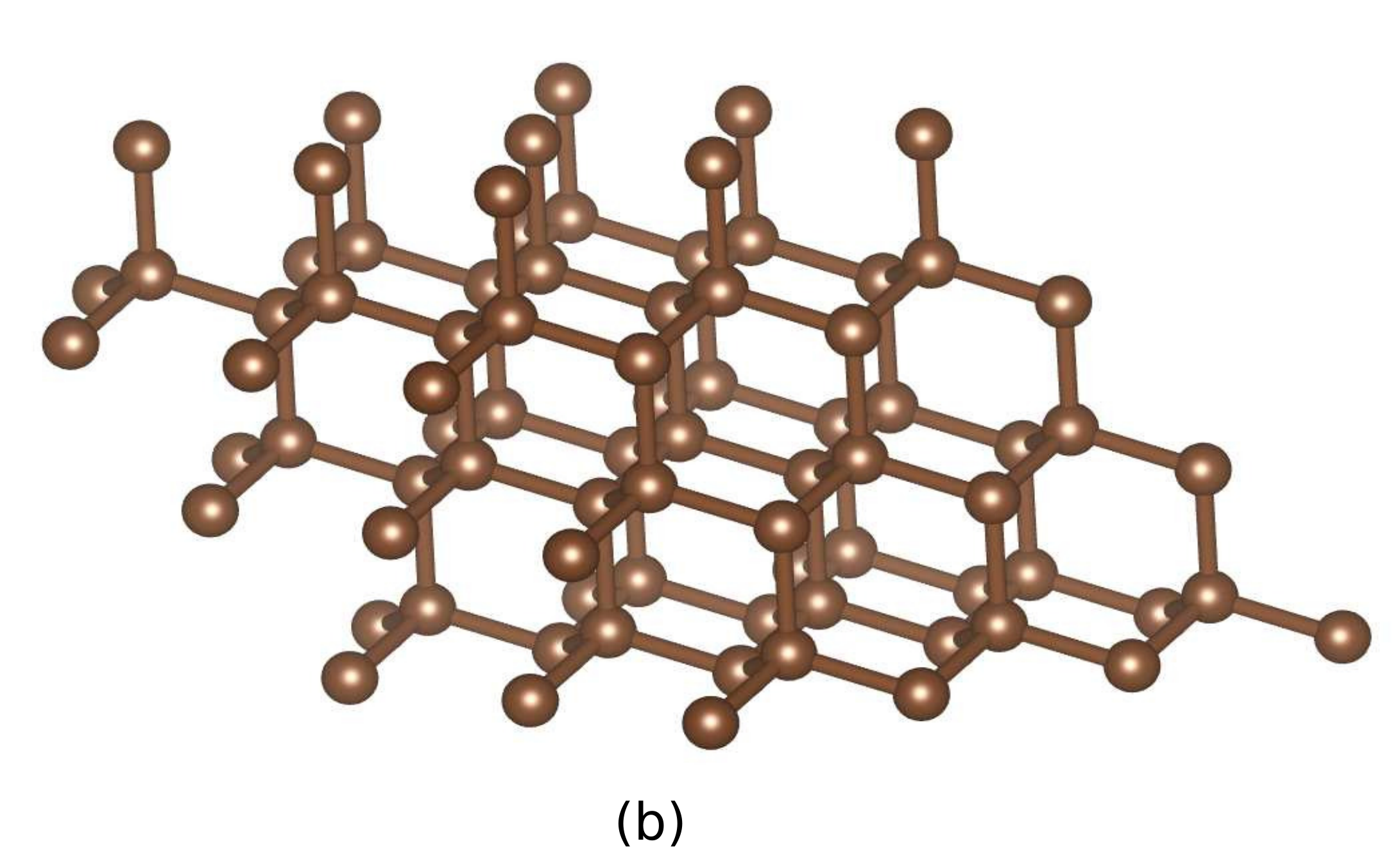}
\caption{The predicted structures for 12 C atoms at 50 GPa. (a) graphite, (b) diamond.}
\label{fig:C}
\end{figure}
\subsection{Metallic systems}
Tests have been performed on metallic systems, including Pd, Ta, Au, Y, and Nb. The known stable structures of these metals were found in the first generation with symmetry constraints (not listed in Table~\ref{diffstr}). For Ta metal, {\sc Muse} found an alternative metastable structure, \textit{Pnma}-Ta,~\cite{Liu2013} which is potentially to be the structure to which bcc Ta transits before melting under high pressure. Y is a metal at ambient conditions. Under compression, it will exhibit superconducting properties according to recent report.~\cite{Chen2012} At 140 GPa, {\sc Muse} also found the superconducting $Fddd$ phase recently reported by Chen \textit{et al.}~\cite{Chen2012} Apart from the $Fddd$ phase, {\sc Muse} also found other energetically competitive structures, such as the $C2/c$ and $P2_1/c$ structures which have not been reported before.

\subsection{Covalent systems}
The covalent systems including C, SiC, and LiBC were also fully tested. For 12 C atoms, {\sc Muse} successfully reproduced the diamond-C and graphite-C at 50 GPa in the same search (see Fig.~\ref{fig:C}). The diamond phase is the most stable one. Other metastable structures were also found, such as $Cmmm$, $Fmmm$, $R\bar3m$, $C2/m$, and $P\bar62m$. The $Fm\bar3m$ structure of 20-atom SiC was easily reproduced in the first generation with symmetry constraints (not listed in Table~\ref{diffstr}). {\sc Muse} also searched many other metastable structures, including $P6_3mc$, $Ccm2_1$, $C2/m$, $Cm$, and $P3m1$. The enthalpy of $P6_3mc$ structure is only 3.1 meV/atom higher than that of $Fm\bar3m$ at 10 GPa. LiBC is an intermetallic compound which also shows strong in-plane covalent bonding. {\sc Muse} only used 4.5 generations to find the known $P6_3/mmc$ structure for the 12-atom system.

\begin{figure}[htp]
\centering
\includegraphics[scale=0.70]{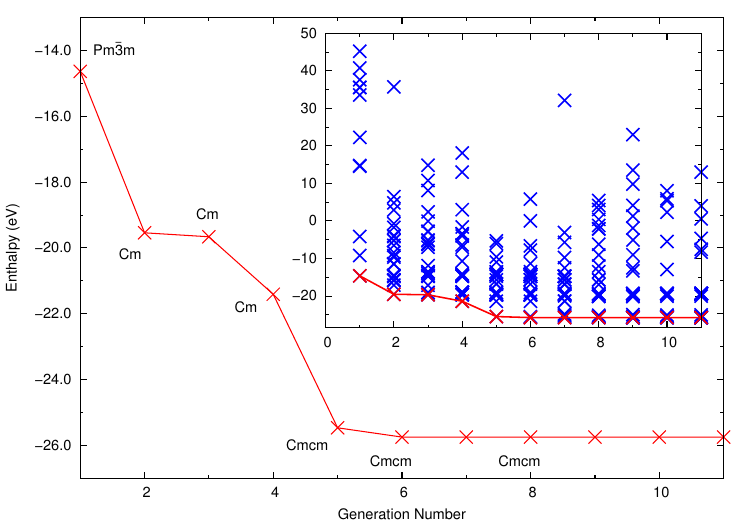}
\caption{The self-adaptive evolution of MgSiO$_3$ structures at 130 GPa. The cell contains 20 atoms. The well-known post-perovskite structure ($Cmcm$) of MgSiO$_3$ was found in the fifth generation. The inset shows all the structures searched.}
\label{fig:mgsio3}
\end{figure}

\subsection{Ionic systems}
NaCl is a typical ionic system. With the help of symmetry constraints on the first generation, {\sc Muse} found its stable $Fm\bar3m$ structure in the first generation using 20 atoms (not listed in Table~\ref{diffstr}). For the tested MgSiO$_3$ system, there were 20 atoms in the unit cell. With so many atoms in the cell, {\sc Muse} found the well-known post-perovskite structure only in seven generations (Table~\ref{diffstr}). Fig.~\ref{fig:mgsio3}  shows the  evolution of MgSiO$_3$ structures at 130 GPa. {\sc Muse} also produced many metastable ones.

\begin{table}
\caption{The efficiency and success rate comparison of each individual algorithm with their combination (MAC). The test system is Al$_2$O$_3$ with 10 atoms. The meanings of the parameters are same as Table~\ref{diffstr}. EA means evolutionary algorithm with mutation and crossover operators. EA2 means EA plus the slip and twist operators.}
\label{comp}
\begin{tabular}{cccccccc}
\hline\hline
	Algorithm & $N_\mathrm{atom}$ & P$_{\mathrm{size}}$ & N$_\mathrm{run}$ &N$_{\mathrm{opt}}$ & Mean N$_{\mathrm{best}}$& \% Success\\
\hline
	SA & 10  & 20 & 20 &734& 1.84 & 95.0\\
	BH & 10  & 20 & 20 & 726 & 1.82 & 85.0\\
	EA & 10  & 20 & 20 & 1588 & 3.97 & 100.0 \\
    EA2 & 10  & 20 & 20 & 774 & 1.94 & 100.0 \\
	MAC & 10 & 20 & 20 & 788 & 1.97 & 100.0\\
\hline\hline
\end{tabular}
\end{table}
\begin{table}
\caption{The influence of the self-adaptive scheme on the efficiency of MAC search. The test system is LiBC with 12 atoms. The meanings of the parameters are the same as Table~\ref{diffstr}.}
\label{compsa}
\begin{tabular}{cccccccc}
\hline\hline
	If self-adaptive&  $N_\mathrm{atom}$ & P$_{\mathrm{size}}$ &N$_\mathrm{run}$ & N$_{\mathrm{opt}}$ & Mean N$_{\mathrm{best}}$& \% Success \\
\hline
	No & 12  & 20 & 3 & 320 & 5.33 & 100.0\\
	Yes & 12 & 20 & 3 & 270 & 4.50 & 100.0\\
\hline\hline
\end{tabular}
\end{table}
\subsection{Efficiency of MAC search}
To further study the efficiency and success rate of MAC search, I performed intensive tests on the Al$_2$O$_3$ system containing 10 atoms at 0 GPa. In detail, twenty searches were performed for each individual algorithm and their combination. The random structures in the first generation were generated with symmetry constraints. The statistical data were achieved by averaging all the test data (see Table~\ref{comp}).  Mean N$_{\mathrm{best}}$ is the averaged generation number to find the lowest-energy structure. It is calculated from $\frac{\mathrm{N}_\mathrm{opt}}{\mathrm{N}_\mathrm{run}\cdot \mathrm{P}_\mathrm{size}}$, where N$_{\mathrm{opt}}$ is the total number of optimized structures, P$_{\mathrm{size}}$ is the population size, and N$_\mathrm{run}$ is the number of searches. We note the SA and BH algorithms have lower success rates. The EA has 100\% success rate but slightly lower efficiency. As we expected, the MAC search has 100\% success rate and higher efficiency. For the newly introduced operators slip and twist, from the comparison with EA  we note they enhance the search ability of {\sc Muse} reasonably (EA2 data in Table~\ref{comp}). In the MAC search, the selection of operators is up to the success rates of operators in the last generation. In other words, the evolution of crystal population is self-adaptive in the manner of breeding offspring. To test such a scheme, I also performed tests on LiBC system with 12 atoms. Tests show the self-adaptive scheme abviously improves the MAC search efficiency (Table~\ref{compsa}). From these tests, we see that the efficiency of {\sc Muse} is greatly improved with the help of the MAC algorithm, the two newly introduced operators, and the self-adaptive scheme.

\section{Conclusion}
\label{disconcl}
In conclusion, I detailed the implementation of MAC crystal structure prediction technique and the testing of the MAC performance, the two new operators, and the self-adaptive scheme. Tests show the multi-algorithm collaborative method is more efficient than individual one. With the help of two new variation operators, slip and twist, the search ability of {\sc Muse} are enhanced. In order to further increase the search efficiency of {\sc Muse}, I increased the number of variation operators to ten. More importantly, I also introduced the competition scheme among the ten variation operators to realize the self-adaptive evolution of crystal population. The symmetry constraints in the first generation, the MAC algorithm, ten variation operators, and the self-adaptive evolution are all key to improving the efficiency of CSP in {\sc Muse}. Present tests show that {\sc Muse} has very high efficiency and 100\% success rate.

\section{Acknowledgements}
The research was supported by the National Natural Science Foundation of China (No. 11104127), the Henan Research Program of Basic and Frontier Technology (No. 102300410213), and the Science Research Scheme of Henan Education Department (No. 2011A140019).


\bibliographystyle{elsarticle-num}

\begin{thebibliography}{10}
\expandafter\ifx\csname url\endcsname\relax
  \def\url#1{\texttt{#1}}\fi
\expandafter\ifx\csname urlprefix\endcsname\relax\def\urlprefix{URL }\fi
\expandafter\ifx\csname href\endcsname\relax
  \def\href#1#2{#2} \def\path#1{#1}\fi

\bibitem{Oganov2006-3}
C.~W. Glass, A.~R. Oganov, N.~Hansen, Comput. Phys. Commun. 175 (2006) 713.

\bibitem{Lonie2011}
D.~C. Lonie, E.~Zurek, Comput. Phys. Commun. 182 (2011) 372.

\bibitem{Maddox1988}
J.~Maddox, Nature 335 (1988) 201.

\bibitem{Martonak2003}
R.~Marto\ifmmode~\check{n}\else \v{n}\fi{}\'ak, A.~Laio, M.~Parrinello, Phys.
  Rev. Lett. 90 (2003) 075503.

\bibitem{Laio2002}
A.~Laio, M.~Parrinello, Proc. Natl. Acad. Sci. 99 (2002) 12562.

\bibitem{Wales1997}
D.~J. Wales, J.~P.~K. Doye, J. Phys. Chem A 101 (1997) 5111.

\bibitem{Goedecker2004}
S.~Goedecker, J. Chem. Phys. 120 (2004) 9911.

\bibitem{Amsler2010}
M.~Amsler, S.~Goedecker, J. Chem. Phys. 133 (2010) 224104.

\bibitem{Oganov2010}
A.~R. Oganov~(Ed.), Modern Methods of Crystal Structure Prediction, Wiley-VCH,
  2010.

\bibitem{evo2013}
S.~Bahmann, J.~Kortus, Comput. Phys. Commun. 184 (2013) 1618.

\bibitem{Wang2010}
Y.~Wang, J.~Lv, L.~Zhu, Y.~Ma, Phys. Rev. B 82 (2010) 094116.

\bibitem{Wang2012}
Y.~Wang, J.~Lv, L.~Zhu, Y.~Ma, Comput. Phys. Commun. 183 (2012) 2063.

\bibitem{Lv2012}
J.~Lv, Y.~Wang, L.~Zhu, M.~Y., J. Chem. Phys. 137 (2012) 084104.

\bibitem{Pickard2011}
C.~J. Pickard, R.~J. Needs, J. Phys.: Condens. Matter 23 (2011) 053201.

\bibitem{Oganov2005}
A.~R. Oganov, R.~Marto\ifmmode~\check{n}\else \v{n}\fi{}\'ak, A.~Laio,
  P.~Raiteri, M.~Parrinello, Nature 438 (2005) 1142.

\bibitem{Ishikawa2006}
T.~Ishikawa, H.~Nagara, K.~Kusakabe, N.~Suzuki,
  Phys. Rev. Lett. 96 (2006) 095502.

\bibitem{Amsler2012a}
M.~Amsler, J.~A. Flores-Livas, L.~Lehtovaara, F.~Balima, S.~A. Ghasemi, D.~Machon, S.~Pailh\`es, A.~Willand, D.~Caliste, S.~Botti, A.~San~Miguel, S.~Goedecker, M.~A.~L. Marques, Phys. Rev. Lett. 108 (2012) 065501.

\bibitem{Flores-Livas2012}
J.~A. Flores-Livas, M.~Amsler, T.~J. Lenosky, L.~Lehtovaara, S.~Botti, M.~A.~L. Marques, S.~Goedecker, Phys. Rev. Lett. 108 (2012) 117004.

\bibitem{Amsler2012}
M.~Amsler, J.~A. Flores-Livas, T.~D. Huan, S.~Botti, M.~A.~L. Marques, S.~Goedecker, Phys. Rev. Lett. 108 (2012) 205505.

\bibitem{Huan2013}
T.~D. Huan, M.~Amsler, M.~A.~L. Marques, S.~Botti, A.~Willand, S.~Goedecker, Phys. Rev. Lett. 110 (2013) 135502.

\bibitem{Oganov2009}
A.~R. Oganov, J.~Chen, C.~Gatti, Y.~Ma, Y.~Ma, C.~W. Glass, Z.~Liu, T.~Yu,
  O.~O. Kurakevych, V.~L. Solozhenko, Nature 457 (2009) 863.

\bibitem{Ma2009}
Y.~Ma, M.~Eremets, A.~R. Oganov, Y.~Xie, I.~Trojan, S.~Medvedev, A.~O. Lyakhov,
  M.~Valle, V.~Prakapenka, Nature 458 (2009) 182.

\bibitem{Gao2008}
G.~Gao, A.~Oganov, A.~Bergara, M.~Martinez-Canalez, T.~Cui, T.~Iitaka, Y.~Ma,
  G.~Zou, Phys. Rev. Lett. 101 (2008) 107002.

\bibitem{Baettig2011}
P.~Baettig, E.~Zurek, Phys. Rev. Lett. 106
  (2011) 237002.

\bibitem{Hermann2012}
A.~Hermann, B.~L. Ivanov, N.~W. Ashcroft, R.~Hoffmann, Phys. Rev. B 86 (2012)
  014104.

\bibitem{Zhu2011}
L.~Zhu, H.~Wang, Y.~Wang, J.~Lv, Y.~Ma, Q.~Cui, Y.~Ma, G.~Zou, Phys. Rev. Lett. 106 (2011) 145501.

\bibitem{Lv2011}
J.~Lv, Y.~Wang, L.~Zhu, Y.~Ma, Phys. Rev. Lett. 106 (2011) 015503.

\bibitem{Wang2011}
Y.~Wang, H.~Liu, J.~Lv, L.~Zhu, H.~Wang, Y.~Ma, Nature Communications 2 (2011)
  563.

\bibitem{Zhu2012}
L.~Zhu, Z.~Wang, Y.~Wang, G.~Zou, H.-k. Mao, Y.~Ma, Proc. Natl. Acad. Sci. 109
  (2012) 751.

\bibitem{Pickard2006}
C.~J. Pickard, R.~J. Needs, Phys. Rev. Lett. 97 (2006) 045504.

\bibitem{Pickard2010}
C.~J. Pickard, R.~J. Needs, Nat. Mater. 9 (2010) 624.

\bibitem{Metropolis1953}
N.~Metropolis, A.~Rosenbluth, M.~Rosenbluth, A.~Teller, E.~Teller, J. Chem.
  Phys. 21 (1953) 1087.

\bibitem{Kresse94}
G.~Kresse, J.~Hafner, Phys. Rev. B 49 (1994) 14251.

\bibitem{Kresse96}
G.~Kresse, J.~Furthm\"{u}ller, Comput. Mater. Sci. 6 (1996) 15.

\bibitem{Soler2002}
J.~M. Soler, E.~Artacho, J.~D. Gale, A.~Garc{\'{i}}a, J.~Junquera,
  P.~Ordej{\'{o}}n, D.~S{\'{a}}nchez-Portal, J. Phys.: Condens. Matter 14
  (2002) 2745.

\bibitem{pwscfcite}
P.~Giannozzi, S.~Baroni, N.~Bonini, M.~Calandra, R.~Car, C.~Cavazzoni,
  D.~Ceresoli, G.~L. Chiarotti, M.~Cococcioni, I.~Dabo, A.~D. Corso, S.~D.
  Gironcoli, S.~Fabris, G.~Fratesi, R.~Gebauer, U.~Gerstmann, C.~Gougoussis,
  A.~Kokalj, M.~Lazzeri, L.~Martin-Samos, N.~Marzari, F.~Mauri, R.~Mazzarello,
  S.~Paolini, A.~Pasquarello, L.~Paulatto, C.~Sbraccia, S.~Scandolo,
  G.~Sclauzero, A.~P. Seitsonen, A.~Smogunov, P.~Umari, R.~M. Wentzcovitch, J.
  Phys. Condens. Matter 21 (2009) 395502.

\bibitem{lammps}
http://lammps.sandia.gov/index.html.

\bibitem{Oganov2006-2}
A.~R. Oganov, C.~W. Glass, J. Chem. Phys. 124 (2006) 244704.

\bibitem{Wales1999}
D.~J. Wales, H.~A. Scheraga, Science 285 (1999) 1368.

\bibitem{Akizuki1981}
M.~Akizuki, Am. Mineral. 66 (1981) 1006.

\bibitem{Mitchell1999}
T.~Mitchell, A.~Misra, Mater. Sci. Eng. A 261 (1999) 106.

\bibitem{Saito1995}
K.~Saito, M.~Asahina, Y.~Yamamura, I.~Ikemoto, J. Phys.: Condens. Matter 7
  (1995) 8919.

\bibitem{Knowles2012}
T.~P.~J. Knowles, A.~De~Simone, A.~W. Fitzpatrick, A.~Baldwin, S.~Meehan,
  L.~Rajah, M.~Vendruscolo, M.~E. Welland, C.~M. Dobson, E.~M. Terentjev, Phys.
  Rev. Lett. 109 (2012) 158101.

\bibitem{Moon2013}
P.~Moon, M.~Koshino, Phys. Rev. B 87 (2013) 205404.

\bibitem{McMahon2001}
M.~McMahon, R.~Nelmes, S.~Rekhi, Phys. Rev. Lett. 87 (2001) 255502.

\bibitem{Nelmes2002}
R.~Nelmes, M.~McMahon, J.~Loveday, S.~Rekhi, Phys. Rev. Lett. 88 (2002) 155503.

\bibitem{Degtyareva2004}
O.~Degtyareva, M.~McMahon, D.~Allan, R.~Nelmes, Phys. Rev. Lett. 93 (2004)
  205502.

\bibitem{Deaven1995}
D.~Deaven, K.~Ho, Phys. Rev. Lett. 75 (1995) 288.

\bibitem{Woodley2007}
S.~Woodley, Phys. Chem. Chem. Phys. 9 (2007) 1070.

\bibitem{Assadollahzadeh2008}
B.~Assadollahzadeh, P.~Bunker, P.~Schwerdtfeger, Chem. Phys. Lett. 451 (2008)
  262.

\bibitem{Abraham2006}
N.~L. Abraham, M.~I.~J. Probert, Phys. Rev. B 73 (2006) 224104.

\bibitem{Trimarchi2007}
G.~Trimarchi, A.~Zunger, Phys. Rev. B 75 (2007) 104113.

\bibitem{Hooper2009}
J.~Hooper, A.~Hu, F.~Zhang, T.~K. Woo, Phys. Rev. B 80 (2009) 104117.

\bibitem{Yeh1992}
C.-Y. Yeh, Z.~W. Lu, S.~Froyen, A.~Zunger, Phys. Rev. B 46 (1992) 10086.

\bibitem{Burdett1987}
J.~K. Burdett, T.~Hughbanks, G.~J. Miller, J.~W. Richardson, J.~V. Smith, J.
  Am. Chem. Soc. 109 (1987) 3639.

\bibitem{Jorgensen1986}
J.-E. Jorgensen, J.~D. Jorgensen, B.~Batlogg, J.~P. Remeika, J.~D. Axe, Phys. Rev. B 33 (1986) 4793.

\bibitem{Pauling1925}
L.~Pauling, S.~B. Hendricks, J. Am. Chem. Soc. 47 (1925) 781.

\bibitem{Fogg2006}
A.~M. Fogg, J.~Meldrum, G.~R. Darling, J.~B. Claridge, M.~J. Rosseinsky, J. Am.
  Chem. Soc. 128 (2006) 10043.

\bibitem{Ono2006}
S.~Ono, T.~Kikegawa, , Y.~Ohishi, Am. Mineral. 91 (2006) 475.

\bibitem{Perdew1996}
J.~P. Perdew, K.~Burke, M.~Ernzerhof, Phys. Rev. Lett. 77 (1996) 3865.

\bibitem{Blochl94}
P.~E. Bl\"{o}chl, Phys. Rev. B 50 (1994) 17953.

\bibitem{Kresse99}
G.~Kresse, D.~Joubert, Phys. Rev. B 59 (1999) 1758.

\bibitem{Liu2013}
Z.~L. Liu, L.~C. Cai, X.~L. Zhang, F.~Xi, J. Appl. Phys. 114 (2013) 073520.

\bibitem{Chen2012}
Y.~Chen, Q.-M. Hu, R.~Yang, Phys. Rev. Lett. 109 (2012) 157004.

\end{thebibliography}

\end{document}